\documentclass[
10pt,
showpacs,preprintnumbers,footinbib,
amsmath,amssymb,
aps,
prl,twocolumn,groupedaddress,superscriptaddress,
showkeys
]{revtex4-1}
\usepackage{epstopdf}
\usepackage{graphicx}
\usepackage{dcolumn}
\usepackage{bm}
\usepackage[colorlinks=true,urlcolor=blue,citecolor=blue,breaklinks=true]{hyperref}
\usepackage{color}
\usepackage{epsfig}
\usepackage{multirow}

\newcommand{\cut}[1]{}

\newcommand{\ket}[1]{\ensuremath{| #1 \rangle}}
\newcommand{\bra}[1]{\ensuremath{\langle #1  |}}

\DeclareMathOperator*{\sumint}{%
\mathchoice%
  {\ooalign{$\displaystyle\sum$\cr\hidewidth$\displaystyle\int$\hidewidth\cr}}
  {\ooalign{\raisebox{.14\height}{\scalebox{.7}{$\textstyle\sum$}}\cr\hidewidth$\textstyle\int$\hidewidth\cr}}
  {\ooalign{\raisebox{.2\height}{\scalebox{.6}{$\scriptstyle\sum$}}\cr$\scriptstyle\int$\cr}}
  {\ooalign{\raisebox{.2\height}{\scalebox{.6}{$\scriptstyle\sum$}}\cr$\scriptstyle\int$\cr}}
}

\begin{document}


\title{Unified description of $^6$Li structure and deuterium-$^4$He dynamics \\ with chiral two- and three-nucleon forces}

\author{Guillaume Hupin}
 \email{ghupin@nd.edu}
 \altaffiliation{Present address:  Department of Physics, University of Notre Dame, Notre Dame, Indiana 46556-5670}
 \affiliation{Lawrence Livermore National Laboratory, P.O. Box 808, L-414, Livermore, California 94551, USA}
\author{Sofia Quaglioni}
 \email{quaglioni1@llnl.gov}
 \affiliation{Lawrence Livermore National Laboratory, P.O. Box 808, L-414, Livermore, California 94551, USA}
\author{Petr Navr\'atil}
 \email{navratil@triumf.ca}
 \affiliation{TRIUMF, 4004 Wesbrook Mall, Vancouver, British Columbia, V6T 2A3, Canada}

\date{\today}

\begin{abstract}
Prototype for the study of weakly bound  projectiles colliding on stable targets, the scattering of deuterium ($d$) on $^4$He ($\alpha$) is an important milestone in the search for a 
fundamental understanding of low-energy reactions. 
At the same time, it is also important for its role in the Big-bang nucleosynthesis of $^6$Li and applications 
in the  characterization of deuterium impurities in materials. 
We present the first unified {\em ab initio} study of the $^6$Li ground state and $d$-$^4$He elastic scattering using two- and three-nucleon forces derived within the framework of chiral effective field theory. The six-nucleon bound-state and scattering observables are calculated by means of the no-core shell model with continuum. 
We analyze the influence of the dynamic polarization of the deuterium and of the chiral three-nucleon force, and examine the role of the continuum degrees of freedom in shaping the low-lying spectrum of $^6$Li. We find that the adopted Hamiltonian 
correctly predicts the binding energy of $^6$Li, yielding an asymptotic $D$- to $S$-state ratio of the $^6$Li wave function in $d+\alpha$ configuration of $-0.027$ in agreement with the value determined from a phase shift analysis of $^6$Li+$^4$He elastic scattering, but overestimates the excitation energy of the first $3^+$ state by $350$ keV. 
The bulk of the computed differential cross section is in good agreement with data. 
\end{abstract}
\maketitle

\paragraph{Introduction.} 

Lithium-6 ($^6$Li) is a weakly-bound stable nucleus that breaks into an $^4$He (or $\alpha$ particle) and a deuteron ($d$) at the excitation energy of $1.4743$ MeV~\cite{Tilley2002a}. 
Until now out of reach of {\em ab initio} (i.e., from first principles) techniques, a complete unified treatment of the bound and continuum properties of this system is desirable to further our understanding of the fundamental interactions among nucleons, but also to 
inform the evaluation of low-energy cross sections relevant to applications. Notable examples are the $^2$H$(\alpha,\gamma)^6$Li radiative capture (responsible for the Big-bang nucleosynthesis of $^6$Li~\cite{Nollett1997,Nollett2001,Hammache2010,Mukhamedzhanov2011,Anders2014}) and the $^2$H$(\alpha,d)^4$He cross section used in the characterization of deuteron concentrations in thin films~\cite{Kellock1993,Quillet1993,Browning2004}.  
Contrary to the lighter nuclei, the structure of the $^6$Li ground state (g.s.)\ -- namely the amount of $D$-state component in its $d+\alpha$ configuration -- is still uncertain~\cite{Tilley2002a}.  
Well known experimentally, the low-lying resonances of $^6$Li  
have been shown to present significant sensitivity to three-nucleon ($3N$) interactions in {\em ab initio} calculations that
treated them as bound states~\cite{Pieper2001,Navratil2003,Pieper2004,Jurgenson2011}. 
However, this approximation is well justified only for the narrow $3^+$ first excited state, and no information about the widths was provided. At the same time, the only {\em ab initio} study of $d$-$^4$He scattering~\cite{Navratil2011} was based on a nucleon-nucleon ($NN$) Hamiltonian and did not take into account the swelling of the $\alpha$ particle due to the interaction with the deuteron. 
  
As demonstrated in a study of the unbound $^7$He nucleus, the {\em ab initio} no-core shell model with continuum (NCSMC)~\cite{Baroni2013} 
is an efficient many-body approach to nuclear bound and scattering states alike. Initially developed to compute nucleon-nucleus collisions starting from a two-body Hamiltonian, this technique has been later extended to include $3N$ forces and successfully applied to make predictions of elastic scattering and recoil of protons off $^4$He~\cite{Hupin2014} and to study continuum and $3N$-force effects on the energy levels of $^9$Be~\cite{Langhammer2014}. We have now developed the NCSMC formalism to describe more challenging deuterium-nucleus collisions, and as a first application, we present in this Letter a study of the $^6$Li ground state and $d$-$^4$He elastic scattering using $NN+3N$ forces 
from chiral effective field theory~\cite{Epelbaum2009,Machleidt2011}.

\paragraph{Approach.}
We cast the microscopic ansatz for the $^6$Li wave function 
in the form of a 
generalized cluster expansion
\begin{align}
\ket{\Psi^{J^\pi \! T}}\! =\! &  \sum_\lambda \!\! c_\lambda \ket{^6 {\rm Li} \, \lambda J^\pi T} 
 \! + \!\sumint_{\nu} \!\! dr \, r^2 
                 \frac{\gamma_{\nu}(r)}{r}
                 {\mathcal{A}}_\nu \ket{\Phi^{J^\pi \! T}_{\nu r}} \,, \label{eq:ansatz}
\end{align}
where $J, \pi$ and $T$ are respectively total angular momentum, parity and isospin, 
$\ket{^6 {\rm Li} \, \lambda J^\pi T}$ represent square-integrable energy eigenstates of the $^6$Li system, and 
\begin{align}
\ket{\Phi^{J^\pi T}_{\nu r}} \!= & 
		\Big[ \!\! \left(
        		\ket{^4 {\rm He} \, \lambda_{\alpha} J_\alpha^{\pi_\alpha}T_\alpha}\ket{^2 {\rm H} \, \lambda_{d} J_{d}^{\pi_d}T_d}
         	\right)^{(sT)} Y_\ell(\hat{r}_{\alpha,d}) \Big]^{(J^{\pi}T)} \nonumber\\ 
	& \times\,\frac{\delta(r-r_{\alpha,d})}{rr_{\alpha,d}} \; 
\label{eq:rgm-state}
\end{align}
are continuous basis states built from a $^4$He and a $^2$H nuclei whose centers of mass are separated by the relative coordinate $\vec r_{\alpha,d}$, 
and that are moving in a $^{2s+1}\ell_J$ partial wave of relative motion.
The translationally-invariant compound, target and projectile states (with energy labels $\lambda, \lambda_\alpha$ and $\lambda_d$, respectively) are all obtained by means of the no-core shell model (NCSM)~\cite{Navratil2000a,Barrett2013} 
using a basis of many-body harmonic oscillator (HO) wave functions with frequency $\hbar\Omega$ and up to $N_{\rm max}$ HO quanta above the lowest energy configuration. The index $\nu$ collects the quantum numbers $\{{}^4{\rm He}\,\lambda_{\alpha} J_{\alpha}^{\pi_{\alpha}}T_{\alpha}; {}^2{\rm H}\,\lambda_{d} J_{d}^{\pi_{d}}T_{d}; s\ell\}$ associated with the continuous basis states of Eq.~(\ref{eq:rgm-state}), and the operator (with $P_{i,j}$ 
exchanging particles $i$ and $j$)
\begin{align}
{\mathcal{A}}_\nu= \frac{1}{\sqrt{15}}\Big(1 - \sum_{i=1}^{4} \sum_{j=5}^{6}P_{i,j}+  \sum_{i<j=1}^{4} P_{i,5} P_{j,6}\Big)\, , \nonumber
\end{align}
ensures its full antisymmetrization. Finally, the unknown discrete coefficients, $c_\lambda$, and continuous amplitudes of relative motion, $\gamma_{\nu}(r)$, are obtained by solving the six-body Schr\"odinger equation in the Hilbert space spanned by the basis states $\ket{^6 {\rm Li} \, \lambda J^\pi T}$ and ${\mathcal A}_{\nu}\ket{\Phi^{J^\pi \! T}_{\nu r}}$~\cite{Baroni2013}
. The bound state and the elements of the scattering matrix are then obtained from matching the solutions of Eq.~(\ref{eq:ansatz}) with the known asymptotic behavior of the 
wave function using an extension of the microscopic $R$-matrix theory~\cite{Hesse1998,Hesse2002}. 

\begin{figure}[t]
\centering
\includegraphics*[width=.9\linewidth]{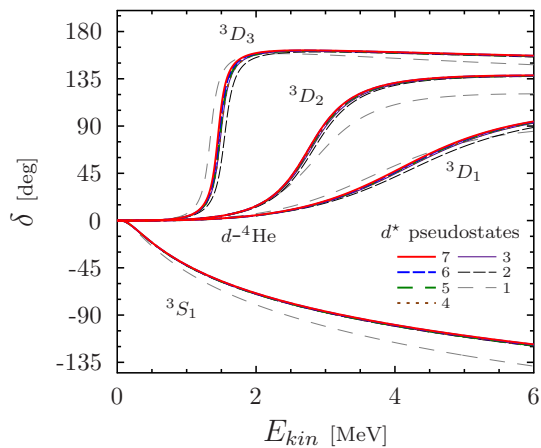}
\caption{(Color online) Computed $d$-$^4$He $S$- and $D$-wave phase shifts at $N_{\rm max}=9$ and $\hbar\Omega=20$ MeV, 
obtained   fifteen square-integrable $^6$Li eigenstates, as a function of the number of $^2$H pseudostates (up to seven) in each of the $^3S_1{-}{}^3D_1$, $^3D_2$ and $^3D_3{-}{}^3G_3$ channels. The two-body part of the SRG-evolved N$^3$LO $NN$ potential ($NN$-only) with $\Lambda=2.0$ fm$^{-1}$ was used.} \label{fig:continuum}
\end{figure}
The deuteron is only bound by $2.224$ MeV. For relative kinetic energies ($E_{kin}$) above this threshold, 
the $d$-$^4$He scattering problem is of a three-body nature (until the breakup of the tightly bound $^4$He, that is). Below, 
the virtual scattering to the energetically closed $^4$He+$p$+$n$ channels 
accounts for the distortion of the projectile. 
Here we address this by discretizing the continuum of $^2$H 
in the $^3S_1{-}{}^3D_1$, $^3D_2$ and $^3D_3{-}{}^3G_3$ channels identified in our earlier study of Ref.~\cite{Navratil2011}.
At the same time, 
fifteen (among which two $1^+$, two $2^+$, and one $3^+$) square-integrable six-body eigenstates of $^6$Li 
also contribute to the description of the deuteron distortion. More importantly, they address the swelling of the $\alpha$ particle~\cite{Hupin2014}, of which we only include the g.s.\ in Eq.~(\ref{eq:rgm-state}). 
The typical convergence behavior of our computed $d$-$^4$He phase shifts with respect to the number of deuteron pseudostates (or $d^\star$, with $E_{d^\star}{>}0$) 
included in Eq.~(\ref{eq:rgm-state}) is shown in  Fig.~\ref{fig:continuum}.
Stable results are found with as little as three deuteron pseudostates per channel. 
This is a strong reduction of the $d^\star$ influence with respect to the more limited study of Ref.~\cite{Navratil2011}, lacking the coupling of square-integrable $^6$Li eigenstates.  
Nonetheless, above the $^2$H breakup threshold, our approach is approximated and a rigorous treatment would require the more complicated task of including three-cluster basis states~\cite{Romero-Redondo2014} 
in the ansazt of Eq.~(\ref{eq:ansatz}). \begin{figure}[t]
\begin{minipage}[c]{.48\linewidth}
\centering
\includegraphics*[width=0.6\linewidth]{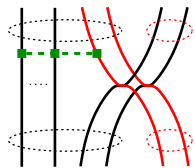}
\end{minipage}
\begin{minipage}[c]{.48\linewidth}
\caption{(Color online)  Diagrammatic representation of one of the $3N$-force matrix elements between basis states of Eq.~(\ref{eq:rgm-state}).
} \label{fig:diagram-NNN}
\end{minipage}
\end{figure}

The treatment of $3N$ forces within the NCSMC formalism to compute deuteron-nucleus collisions involves major technical and computational challenges. 
The first is the derivation and calculation of the matrix elements between the continuous basis states of Eq.~(\ref{eq:rgm-state}) of seven independent $3N$-force terms, five of which  
involve the exchange of  one or two nucleons belonging to the projectile with those of the target. A typical example is 
the diagram of Fig.~\ref{fig:diagram-NNN}, 
which for the present application corresponds to $\bra{\Phi^{J^\pi T}_{\nu^\prime r^\prime}} P_{3,5}P_{4,6}V^{3N}_{123}\ket{\Phi^{J^\pi T}_{\nu r}}$, with $V^{3N}_{123}$ the 
$3N$ interaction among particles $1,2$ and $3$.
To calculate this contribution, we need the four-nucleon density matrix of the target. For $^4$He, this can be precomputed and stored in a factorized form~\cite{Navratil2011,Hupin2013}. 
An additional difficulty 
is represented by the exorbitant number of  input $3N$-force matrix elements (see Fig.~1 of Ref.~\cite{Roth2014}), 
which we have to limit by specifying a maximum  three-nucleon HO model space size $E_{3{\rm max}}$~\cite{Hupin2013}. To minimize the effects of such truncation we included 
$3N$-force matrix elements up to $E_{3{\rm max}}=17$. 
The 
$\langle {}^6{\rm Li} \lambda J^{\pi}T | V^{3N}_{346}| \Phi_{\nu r}^{J^\pi T} \rangle $ and $ \langle {}^6{\rm Li} \lambda J^{\pi}T | V^{3N}_{456}  | \Phi_{\nu r}^{J^\pi T} \rangle $ couplings between discrete and continuous states are 
comparatively less demanding. 

\paragraph{Results.}
\begin{figure}[h]
\centering
\includegraphics*[width=.95\linewidth]{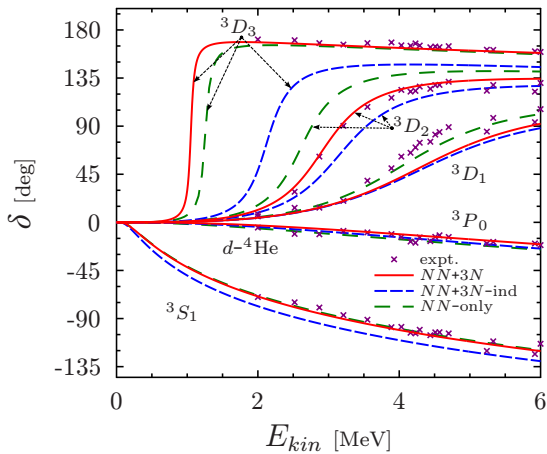}
\caption{(Color online) $S$-, $^{3}P_0$- and $D$-wave $d$-$^4$He phase shifts 
computed with the $NN$-only, $NN$+$3N$-ind and $NN$+$3N$ Hamiltonians (lines) compared to those  extracted from $R$-matrix analyses of data~\cite{Gruebler1975,Jenny1983} (symbols). More details in the text.} \label{fig:phaseshift-interaction}
\end{figure}
We adopt an Hamiltonian based on the chiral N$^3$LO $NN$ interaction of Ref.~\cite{Entem2003} and N$^2$LO $3N$ force of Ref.~\cite{Navratil2007}, constrained to provide an accurate description of the $A{=}2$ and $3$~\cite{Gazit2009} systems. These interactions are additionally softened by means of a unitary transformation that decouples high- and low-momentum components of the Hamiltonian, working within the similarity renormalization group (SRG) method~\cite{Bogner2007,Hergert2007,Jurgenson2009,Roth2011,Roth2014}. To minimize the occurrence of induced four-nucleon forces, 
we work with the SRG momentum scale $\Lambda=2.0$ fm$^{-1}$~\cite{Jurgenson2009,Roth2011,Hupin2013}. All calculations are carried out using the ansatz of Eq.~(\ref{eq:ansatz}) with fifteen discrete eigenstates of the $^6$Li system and continuous $d$-$^4$He(g.s.)\ binary-cluster states with up to seven deuteron pseudostates in the $^3S_1{-}{}^3D_1$, $^3D_2$ and $^3D_3{-}{}^3G_3$ channels. 
Similar to our earlier study of $d$-$^4$He scattering~\cite{Navratil2011} [performed with a softer $NN$ interaction but in a model space spanned only by the continuous basis states of Eq.~(\ref{eq:rgm-state})], we approach convergence for the HO expansions at $N_{\rm max}=11$. 
We adopt the 
HO frequency
of $20$ MeV 
around which the $^6$Li g.s.\ energy calculated within the square-integrable basis of the NCSM becomes nearly insensitive to $\hbar\Omega$~\cite{Jurgenson2011}.  

We start by 
discussing the influence of $3N$ forces -- those induced by the SRG transformation of the $NN$ potential ($NN$+$3N$-ind) as well as those initially present in the chiral Hamiltonian ($NN$+$3N$). In Fig.~\ref{fig:phaseshift-interaction} we compare our computed $d$-$^4$He $S$-, $^{3}P_0$- and $D$-wave phase shifts with those of the $R$-matrix analyses of Refs.~\cite{Gruebler1975,Jenny1983}. The results based on
the two-body part 
of the evolved $NN$ force ($NN$-only) 
resemble those obtained with a softer potential~\cite{Navratil2011}. 
Once the SRG unitary equivalence is restored via the induced $3N$ force, the resonance centroids 
are systematically shifted to higher energies. By contrast, the agreement with data is much improved in the $NN$+$3N$ case and, in particular, the 
splitting between the $^{3}D_{3}$ and $^{3}D_{2}$ partial waves is comparable to the measured one. 
%
\begin{figure}[t]
\vspace{-4mm}
\centering
\includegraphics*[width=.95\linewidth]{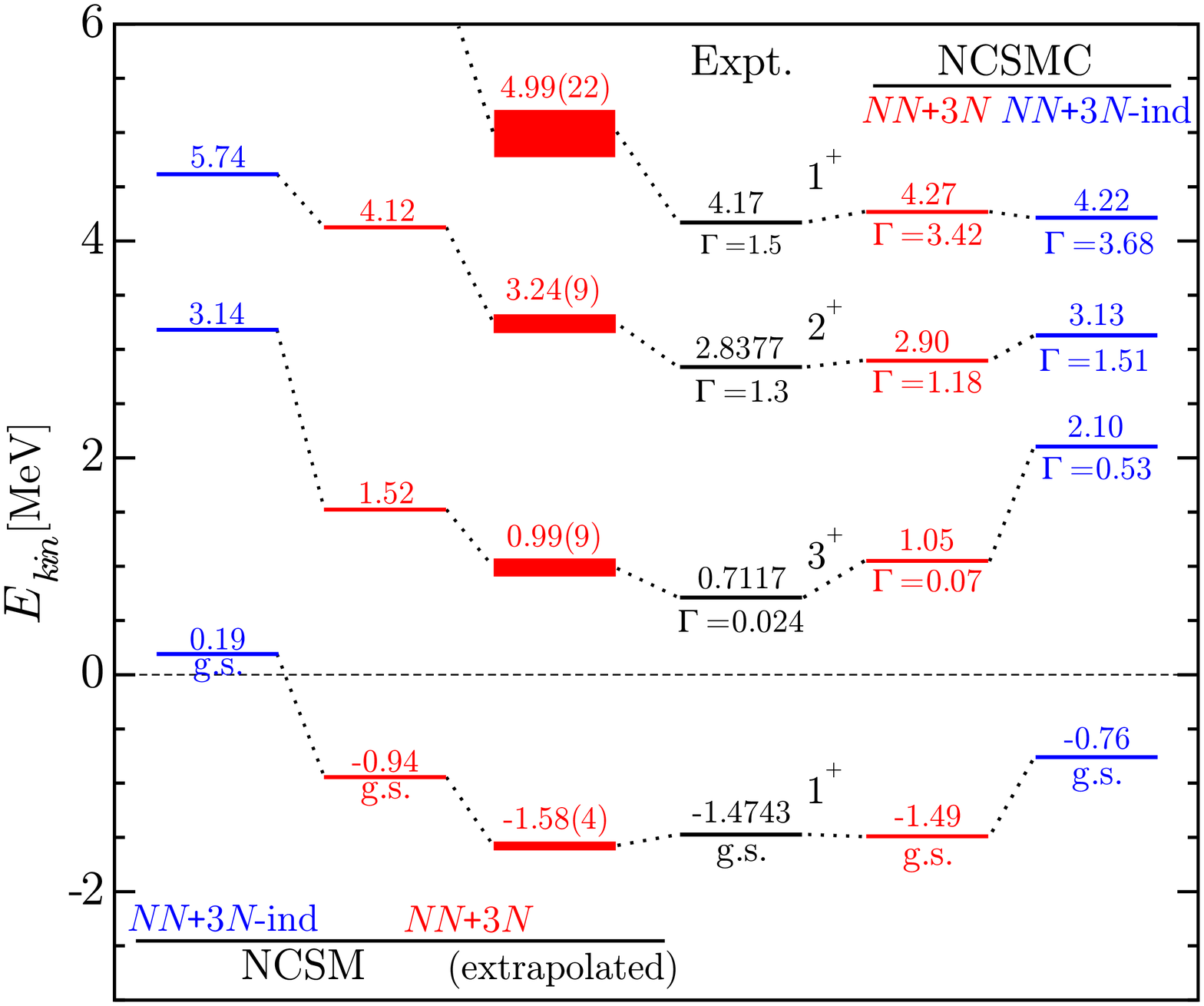}
\caption[]{(Color online)  Ground-state energy and low-lying $^6$Li  positive-parity $T=0$ resonance parameters extracted~\cite{Note1}
from the phase shifts of Fig.~\ref{fig:phaseshift-interaction} (NCSMC) compared to the evaluated centroids and widths (indicated by $\Gamma$) of Ref.~\cite{Tilley2002a} (Expt.). Also shown on the left-hand-side are the best ($N_{\rm max}=12$) and extrapolated~\cite{Note2}
NCSM energy levels. 
The zero energy is set to the respective computed (experimental) $d+^4$He breakup thresholds. 
} \label{fig:spectrum}
\end{figure}
%
%
\begin{table}[b]
\caption[]{Absolute $^6$Li g.s.\ energy, $S$- ($C_0$) and $D$-wave ($C_2$) asymptotic normalization constants and their ratio using the 
$NN+3N$ Hamiltonian compared to experiment. 
Indicated in parenthesis is the $N_{\rm max}$ value of the respective calculation.
The error estimates quoted in the extrapolated ($\infty$) NCSM results include uncertainties due to the SRG evolution of the Hamiltonian and $\hbar\Omega$ dependence~\cite{Jurgenson2011}.
\label{table:absolute-value}}
\begin{ruledtabular}
\renewcommand{\arraystretch}{1.3}
\begin{tabular}{l  c  c  c  c  c  }
Ground-State&$E_{\rm g.s.}$ &  $C_0$ & $C_2$ & $C_2/C_0$ \\
Properties& [MeV]& [fm$^{-1/2}$] & [fm$^{-1/2}$] & \\
\hline
NCSM (10) & -30.84 & $-$ & $-$ & $-$\\
NCSM (12) & -31.52 & $-$ & $-$ & $-$\\
NCSM ($\infty$)~\cite{Note2} & \;\;\,-32.2(3) & $-$ & $-$ & $-$\\
NCSMC (10) & -32.01 & \!\!\!\!\!\!\!2.695 & \!\!\!\!\!\!\!\!\!\!\!-0.074 & \!\!\!\!\!\!\!\!\!\!\!\!\!\!\!\!\!\!\!-0.027\\
Expt.\cite{Tilley2002a,Audi2012,
George1999} & -31.99 & \!\!\!2.91(9) & -0.077(18) & -0.025(6)(10)\\
Expt.~\cite{Blokhintsev1993,Veal1998} & $-$ & $2.93(15)$ & $-$ & \!\!\!\!\!0.0003(9)
\end{tabular}
\end{ruledtabular}
\end{table}

\begin{figure*}
\begin{minipage}[c]{.32\linewidth}
\centering
\includegraphics*[width=.94\linewidth]{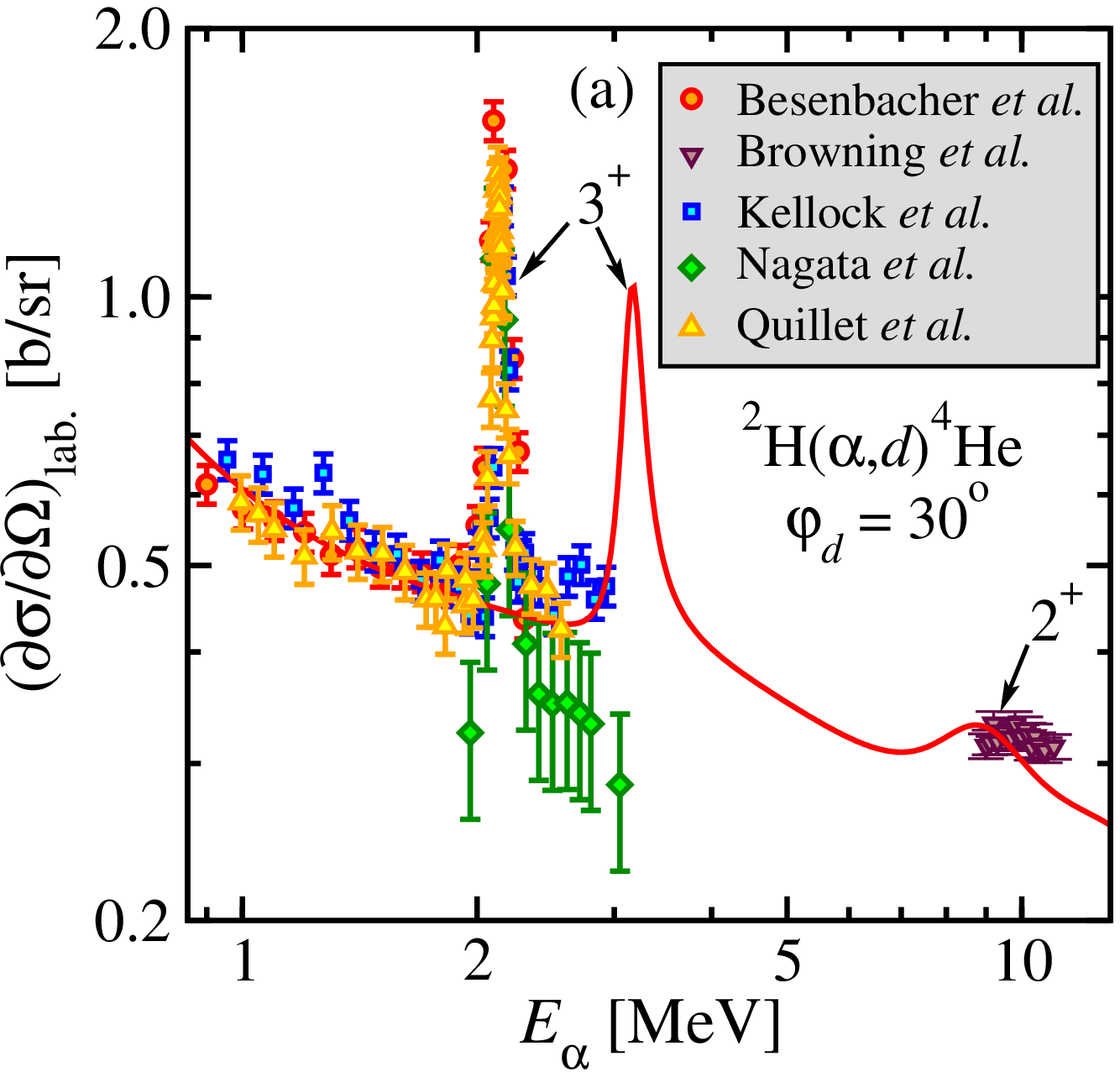}
\end{minipage}
\hfill
\begin{minipage}[c]{.32\linewidth}
\centering
\includegraphics*[width=.94\linewidth]{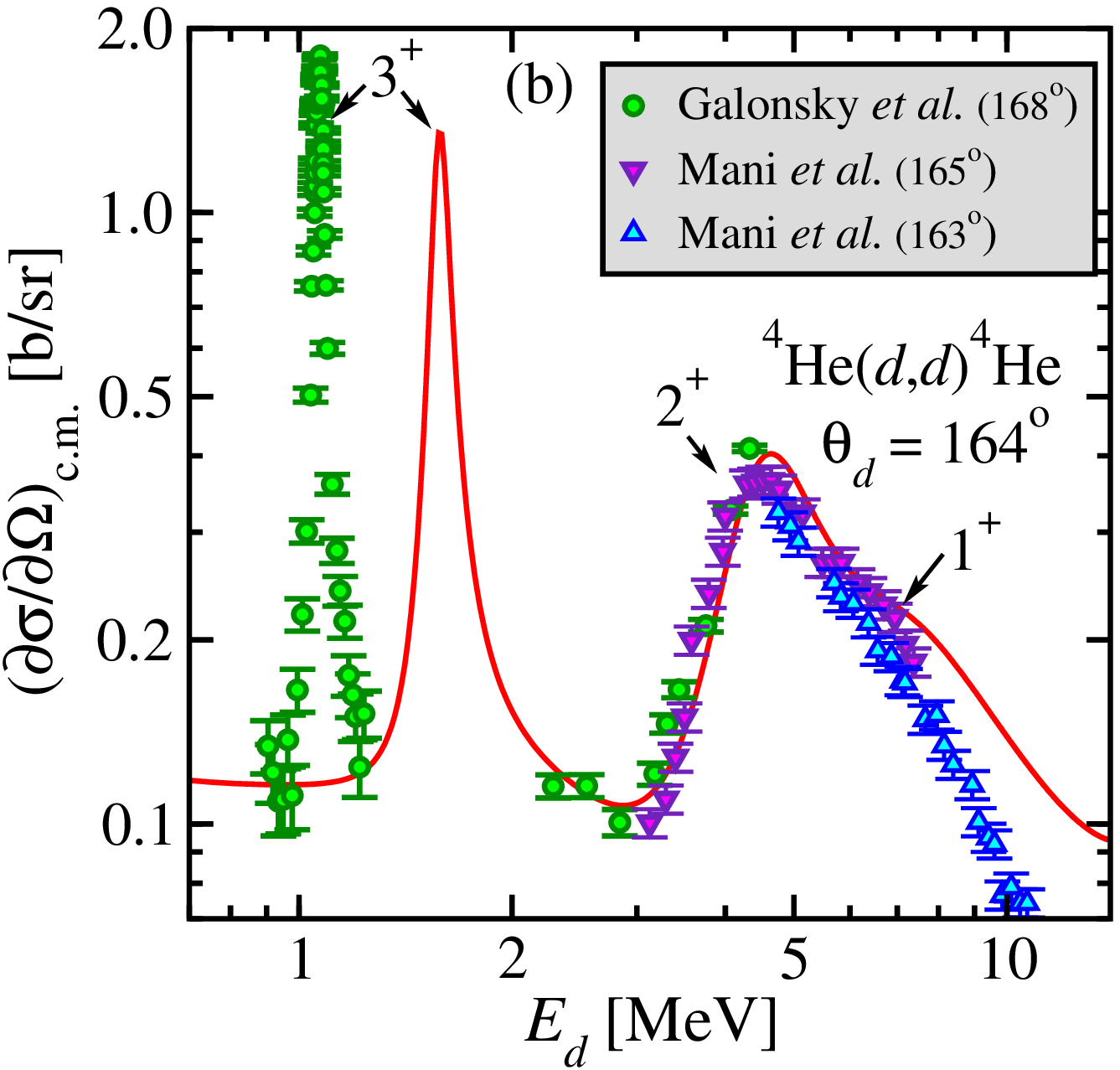}
\end{minipage}
\hfill
\begin{minipage}[c]{.32\linewidth}
\centering
\includegraphics*[width=.94\linewidth]{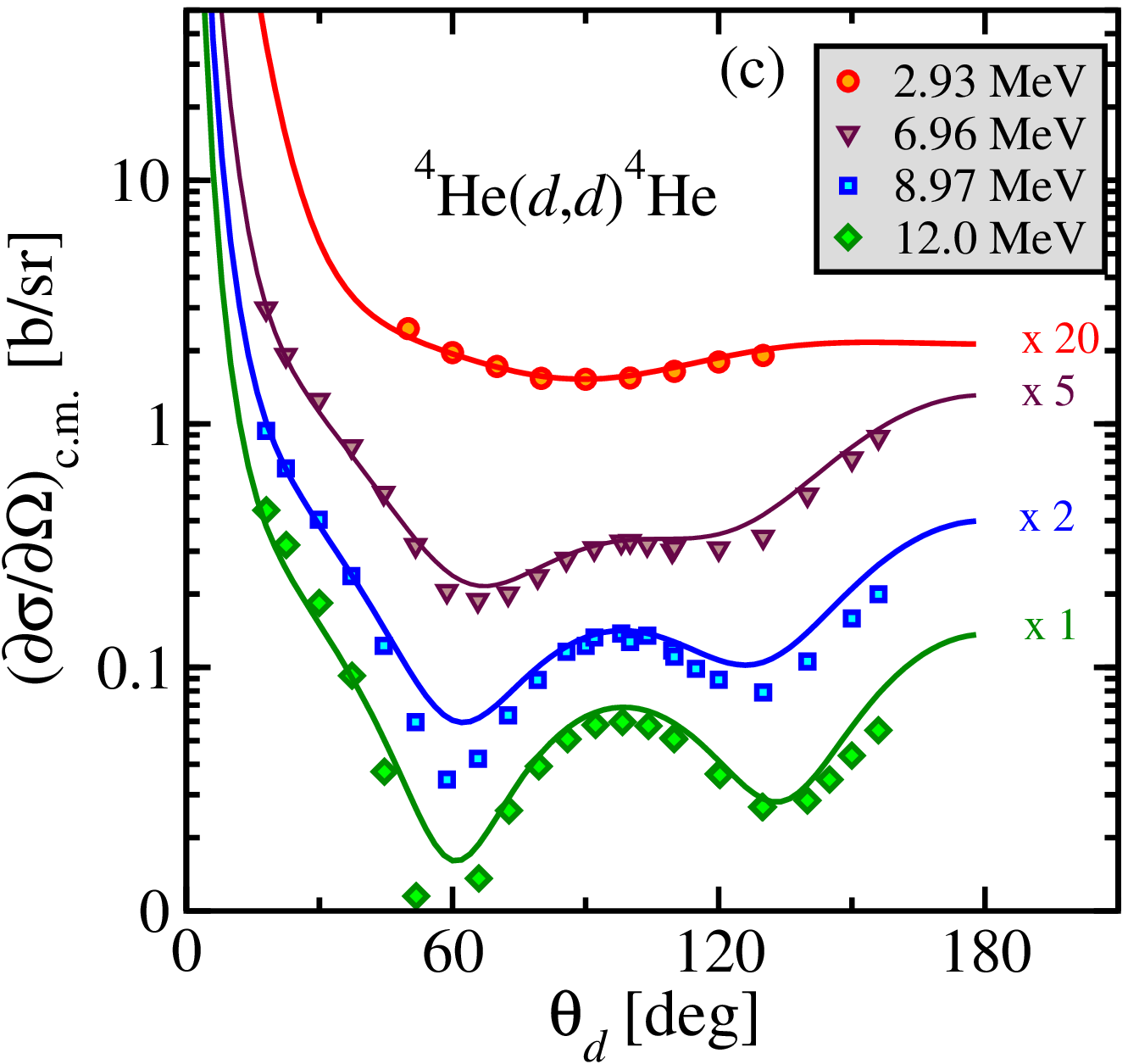}
\end{minipage}
\caption{
(Color online) Computed (a) $^2$H$(\alpha,d)^4$He laboratory-frame and (b) $^4$He$(d,d)^4$He center-of-mass frame angular differential cross sections (lines) using the $NN+3N$ Hamiltonian at the deuteron recoil and backscattered angles of, respectively, $\varphi_d=30^\circ$ and $\theta_d=164^\circ$ as a function of the laboratory helium ($E_\alpha$) and deuteron ($E_d$) incident energies, compared with data (symbols) from Refs.~\cite{Besenbacher1986,Browning2004,Kellock1993,Nagata1985,Quillet1993,Galonsky1955,Mani1968}. In panel (c), calculated (lines) and measured (symbols) center-of-mass angular distributions at $E_d=2.93,6.96, 8.97$~\cite{Jett1971}, and $12$ MeV~\cite{Senhouse1964} are scaled by a factor of $20,5,2$, and $1$, respectively. All positive- and negative-parity partial waves up to $J=3$ were included in the calculations.} \label{fig:Xsection-energy}
\end{figure*}
In Fig.~\ref{fig:spectrum},
the resonance centroids and widths extracted~
\footnote{Centroids $E_R$ and widths $\Gamma$ are obtained, respectively, as the values of $E_{kin}$ for which the first derivative $\delta^\prime(E_{kin})$ of the phase shifts is maximal and $\Gamma{=}2/\delta^\prime(E_R)$.}
from the 
phase shifts of Fig.~\ref{fig:phaseshift-interaction} (shown on the right) are compared with experiment as well as with more traditional  approximated energy levels (shown on the left) obtained within the NCSM by treating the $^6$Li excited states as bound states. 
In terms of excitation energies relative to the g.s., in both calculations 
(i.e.,\ with or without continuum effects) the chiral $3N$ force affects mainly the splitting between the $3^+$ and $2^+$ states, and to a lesser extent the position of the first excited state. 
Sensitivity to the chiral $3N$ force is also seen in the widths of the NCSMC resonances, which tend to become narrower (in closer agreement with experiment) when this force is present in the initial Hamiltonian. Overall, the closest agreement with the observed spectrum is obtained with the $NN$+$3N$ Hamiltonian working within the NCSMC, i.e.\ by including the continuum degrees of freedom. 
Compared to the best ($N_{\rm max}=12$) NCSM values,  all resonances are shifted to lower energies commensurately with their distance from the $d$+$^4$He breakup threshold. For the $3^+$, which is a narrow resonance, the effect is not sufficient to correct for the slight overestimation in excitation energy already observed in the NCSM calculation. This and the ensuing underestimation of the splitting between the $2^+$ and $3^+$ states point to remaining deficiencies in the adopted $3N$ force model, particularly concerning the strength of the spin-orbit interaction. 

The inclusion of the $d$+$^4$He 
states of Eq.~(\ref{eq:rgm-state}) results also in additional binding  for the $1^+$ ground state. This 
stems from a more efficient description of the clusterization of $^6$Li into $d+\alpha$ 
at long distances, which is harder to describe within a finite HO model space, or -- more simply -- from the increased size of the many-body model space. Indeed, as shown in Fig.~\ref{fig:spectrum} and in Table~\ref{table:absolute-value} for the absolute value of the $^6$Li g.s.\ energy, extrapolating to $N_{\rm max}\rightarrow\infty$~
\footnote{Extrapolated values $E_{\infty}$ are obtained from fitting the $N_{\rm max}{=}6$ to $12$ energies at $\hbar\Omega{=}20$ MeV with the function $E(N_{\rm max}){=}E_{\infty}{+}a\,\exp(-b\, N_{\rm max})$.} 
brings the NCSM results in good agreement with the NCSMC, particularly for bound states and narrow resonances. However, while the extrapolation procedure yields comparable energies, only the NCSMC wave functions present the correct  asymptotic, which for the g.s.\ is a Whittaker function. This is essential for the extraction of the asymptotic normalization constants and a future description of the  $^2$H$(\alpha,\gamma)^6$Li radiative capture~\cite{Mukhamedzhanov2011}. The obtained asymptotic $D$- to $S$-state 
ratio 
is not compatible with the near zero value of Ref.~\cite{Veal1998}, but rather is in good agreement with the 
determination of Ref.~\cite{George1999}, stemming from 
an analysis of $^6$Li+$^4$He elastic scattering. 
Further, 
based on the extrapolated NCSM energies, one could erroneously conclude that the measured splitting between $2^+$ and $3^+$ state is reproduced with the $NN+3N$ Hamiltonian.  Conversely, the square-integrable 
$\ket{^6 {\rm Li} \, \lambda\, J^\pi T}$  
components of 
Eq.~(\ref{eq:ansatz}) 
are key to 
achieving an efficient description of the short-range six-body correlations, and compensate for computationally arduous to include $^4$He excited states. 

Next, in Figs.~\ref{fig:Xsection-energy}(a) and \ref{fig:Xsection-energy}(b), respectively, we 
compare the $^2$H$(\alpha,d)^4$He deuteron elastic recoil and $^4$He$(d,d){}^4$He deuteron elastic 
scattering differential cross sections 
computed 
using the $NN$+$3N$ Hamiltonian 
to the measured energy distributions of Refs.~\cite{Besenbacher1986,Browning2004,Kellock1993,Nagata1985,Quillet1993,Galonsky1955,Mani1968}. 
Aside from the position of the  $3^+$ resonance, 
the calculations are in fair agreement with experiment, particularly in the low-energy region of interest for the Big-bang nucleosynthesis of $^6$Li, 
where we reproduce the data of Besenbacher {\em et al.}~\cite{Besenbacher1986} and those of Quillet {\em et al.}~\cite{Quillet1993}. The $500$ keV region below the resonance in Fig.~\ref{fig:Xsection-energy}(a) is also important for material science, where the elastic recoil of deuterium knocked by incident $\alpha$ particles is used to analyze the presence of this element. At higher 
energies, 
near the $2^+$ and $1^+$ resonances, 
the computed 
cross section at the center-of-mass deuteron scattering angle of $\theta_d=164^\circ$ reproduces the data of Galonsky {\em et al.}~\cite{Galonsky1955} and Mani {\em et al.}~\cite{Mani1968}, while we find slight disagreement with the data of Ref.~\cite{Browning2004} 
in the elastic recoil configuration at the laboratory angle of $\varphi_d=30^\circ$. At even higher energies, the measured cross section of  Fig.~\ref{fig:Xsection-energy}(b) lies below the calculated one. 
This is due to the overestimated width of the $1_2^+$ state, which is twice as large as in experiment. The overall good agreement with experiment is also corroborated by Fig.~\ref{fig:Xsection-energy}(c), presenting $^4$He$(d,d){}^4$He angular distributions in the $2.93\le E_d\le12.0$ MeV interval of 
incident energies. In particular, the theoretical curves reproduce the data at $2.93$ and $6.96$ MeV, while some deviations are visible at the two higher energies, in line with our previous discussion. Nevertheless, in general the present results with $3N$ forces provide a much more realistic description of the scattering process than our earlier study of Ref.~\cite{Navratil2011}.  
Finally, we expect that an $N_{\rm max}=13$ calculation (currently out of reach) would not significantly change the present picture,
particularly concerning 
the narrow $3^+$ resonance. Indeed, much as in the case of the g.s.\ energy, here 
the 
NCSMC centroid 
is in good agreement  with the NCSM extrapolated value, $0.99(9)$ MeV. 

\paragraph{Conclusions.}
We presented the first application of the {\em ab initio} NCSMC formalism to the description of deuteron-nucleus dynamics. 
We illustrated the role of the chiral $3N$ force and continuous degrees of freedom in determining the bound-state properties of $^6$Li and $d$-$^4$He elastic scattering observables. The computed g.s.\ energy is in excellent agreement with experiment, and our $d$+$\alpha$ asymptotic normalization constants 
support a non-zero negative ratio of $D$- to $S$-state components for $^6$Li.      
We used data for deuterium backscattering and recoil cross sections 
of interests to ion beam spectroscopy to validate 
our scattering calculations and found a good agreement in particular at low energy. The overestimation by about 350 keV of the position of the $3^+$ resonance is an indication of remaining deficiencies of the nuclear Hamiltonian employed here. This work sets the stage for the first {\em ab initio} study of the $^2$H$(\alpha,\gamma)^6$Li radiative capture, and is a stepping stone in the calculation of the deuterium-tritium fusion with the chiral $NN+3N$ Hamiltonian, currently in progress. 

%
\begin{acknowledgments}
Computing support for this work came from the Lawrence Livermore National Laboratory (LLNL) institutional Computing Grand Challenge
program. It was prepared in part by LLNL under Contract No. DE-AC52-07NA27344. This material is based upon work supported by the U.S.\ Department of Energy, Office of Science, Office of Nuclear Physics, under 
Work Proposal No. SCW1158, 
and by the NSERC Grant No. 401945-2011. TRIUMF receives funding via a contribution through the 
Canadian National Research Council.
\end{acknowledgments}

\bibliographystyle{apsrev4-1}

%

\end{document}